\documentclass{an}
\usepackage{graphicx}
\usepackage{times}
\usepackage{fancyhdr}
\begin{document}

\title{Calibrating models of ultralow-mass stars}

\author{A. Reiners$^{1,2,\star}$}
\institute{$^1$Department of Astronomy, 521 Campbell Hall, University of California, Berkeley, CA 94720-3411, USA\\
$^2$Hamburger Sternwarte, Gojenbergsweg 112, 21029 Hamburg, Germany}

\date{Received; accepted; published online}

\abstract{Evolutionary and atmospheric models have become available
  for young ultralow-mass objects. These models are being used to
  determine fundamental parameters from observational properties. TiO
  bands are used to determine effective temperatures in ultralow-mass
  objects, and together with Na- and K-lines to derive gravities at
  the substellar boundary.  Unfortunately, model calibrations in
  (young) ultralow-mass objects are rare. As a first step towards a
  calibration of synthetic spectral features, I show molecular bands
  of TiO, which is a main opacity source in late M-dwarfs. The TiO
  $\epsilon$-band at 8450\AA\ is systematically too weak. This implies
  that temperatures determined from that band are underestimated, and
  I discuss implications for determining fundamental parameters from
  high resolution spectra.  }

\correspondence{areiners@berkeley.edu\\$^\star$Marie Curie International Outgoing Fellow}

\maketitle

\section{Introduction}

The amount of observational data on young very low mass objects is
growing rapidly. The classification into spectral types according to
major absorption features, and a comparison to spectral energy
distributions can provide a temperature scale that extends down to the
coolest known objects and connects to stars of spectral type M and
hotter (Golimowski et al., 2004).  However, fundamental parameters for
ultralow-mass objects, particularly for very young objects, are only
poorly known and theory is not yet able to reliably predict such
parameters.  A number of discoveries have casted doubt on the
applicability of evolutionary tracks at young ages and low masses
(e.g., Hillenbrand \& White 2004; Close et al. 2005, Reiners et al.
2005) -- a non surprising fact since the models are not expected to
hold at such young ages anyway.

Tests of evolutionary models come from systems where preferably mass
and age are known, i.e. predominantly from binaries in clusters or
star forming regions. Masses of binaries can be measured directly and
provide the most reliable test of theory. A second way to obtain
fundamental parameters without using evolutionary models is to derive
them from comparison to synthetic spectra. In high resolution spectra
($R \ga 20\,000$), a number of sensitive tracers can be found which
are sensitive to temperature and gravity.  Mohanty et al. (2004a,
2004b) have recently derived the fundamental parameters temperature,
gravity, radius and mass from high resolution spectra for a number of
members of the UpSco association (5\,Myr). They compared their results
to evolutionary tracks from Baraffe et al. (1998) and Chabrier,
Baraffe \& Hauschildt (2000), and found significant disagreement
between their observations and evolutionary models.

Stellar parameters derived from comparison to synthetic spectra are
independent of evolutionary models, but completely depend on the
spectral synthesis involving atmospheric (temperature) structure and
molecular data. In the following, currently available synthetic high
resolution spectra generated with the PHOENIX code shall be compared
to spectra of stars with known atmospheric parameters $T_{\rm eff}$,
log\,$g$ and metallicity. I will show that some tracers of the current
models do not yield correct temperatures and I will discuss the
implications for derived temperatures and gravities.

\section{Temperature and gravity from high resolution spectra}

\begin{figure*}
  \mbox{
    \resizebox{.25\hsize}{!}{\includegraphics[]{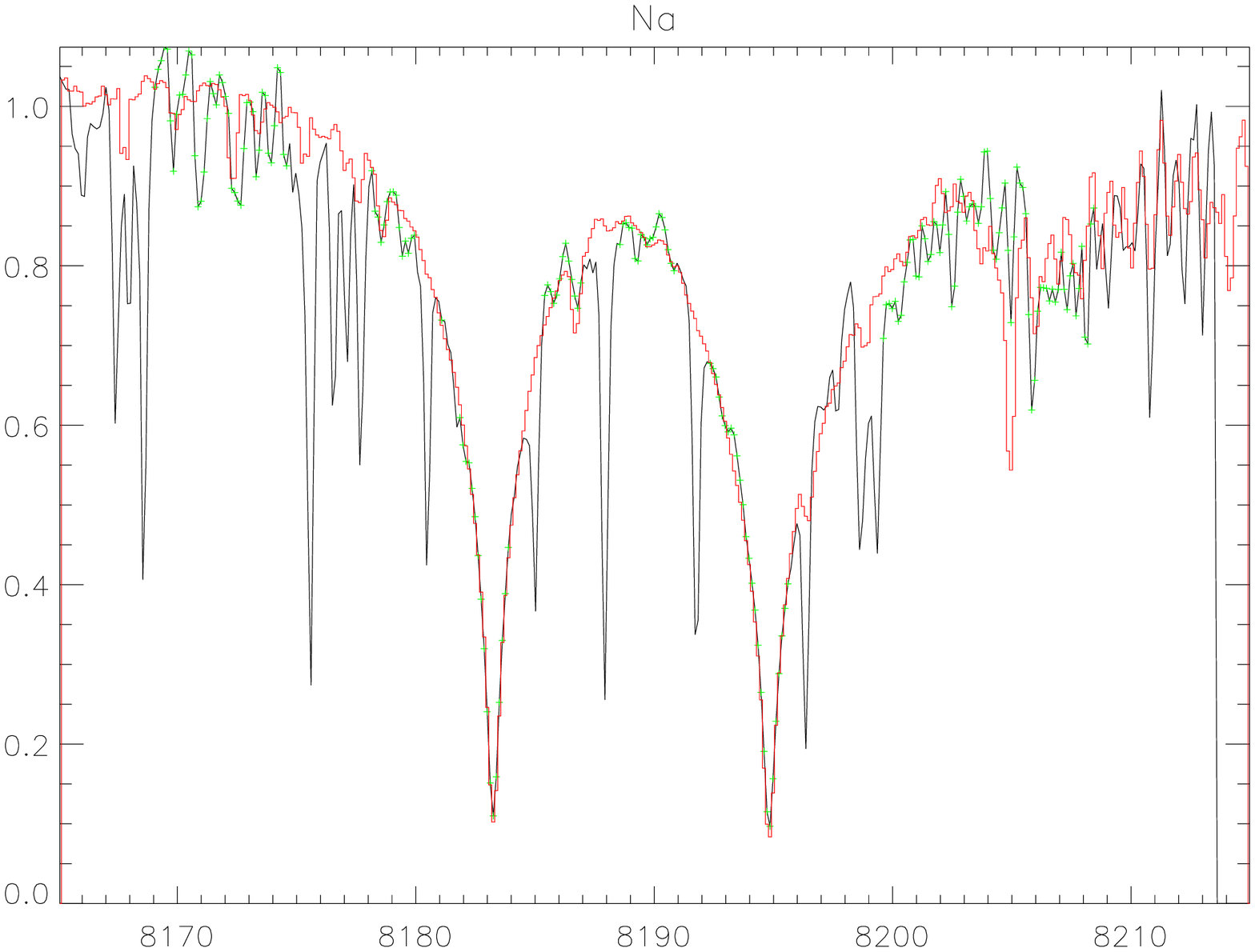}}
    \resizebox{.25\hsize}{!}{\includegraphics[]{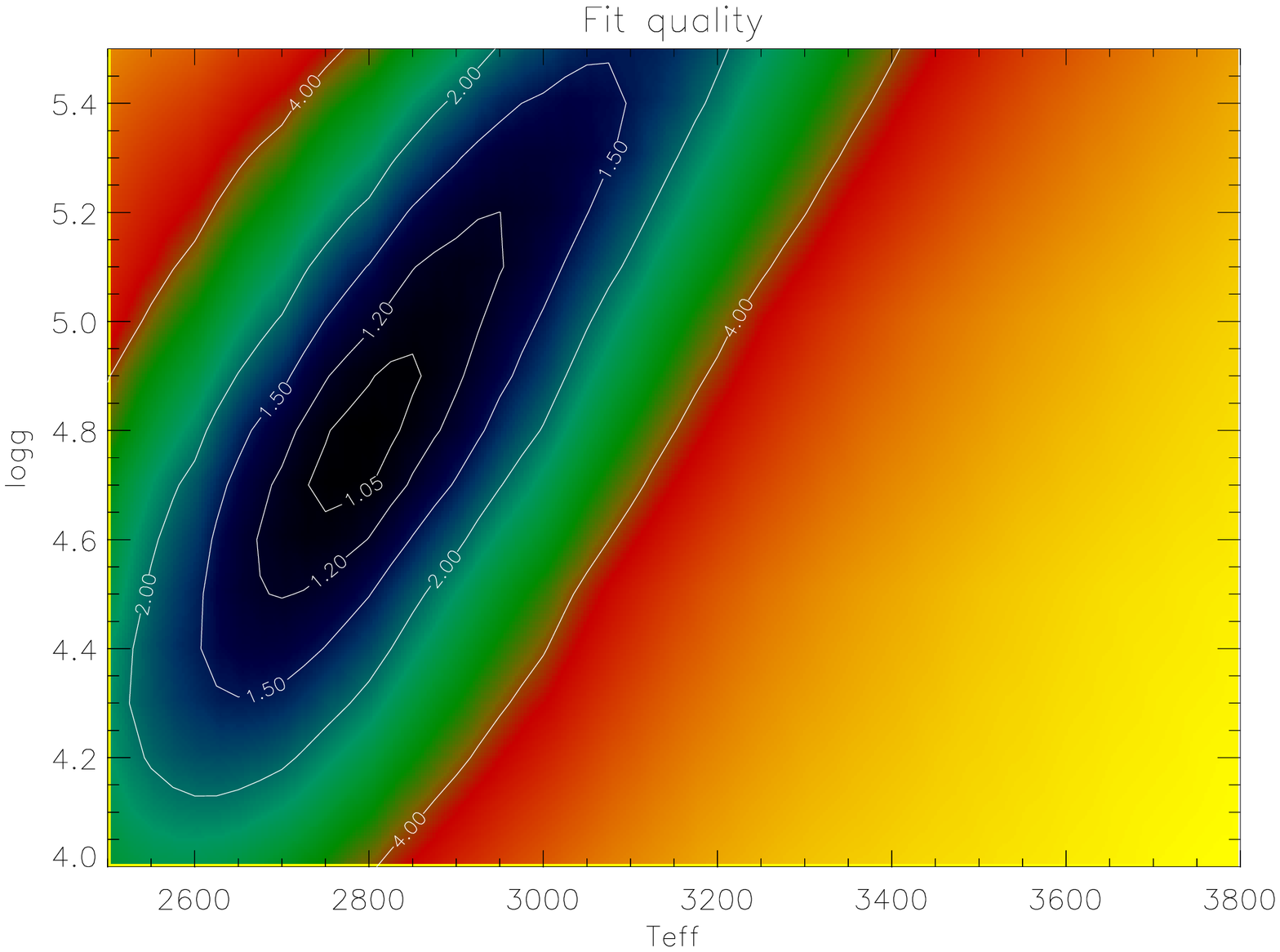}}
    \resizebox{.25\hsize}{!}{\includegraphics[]{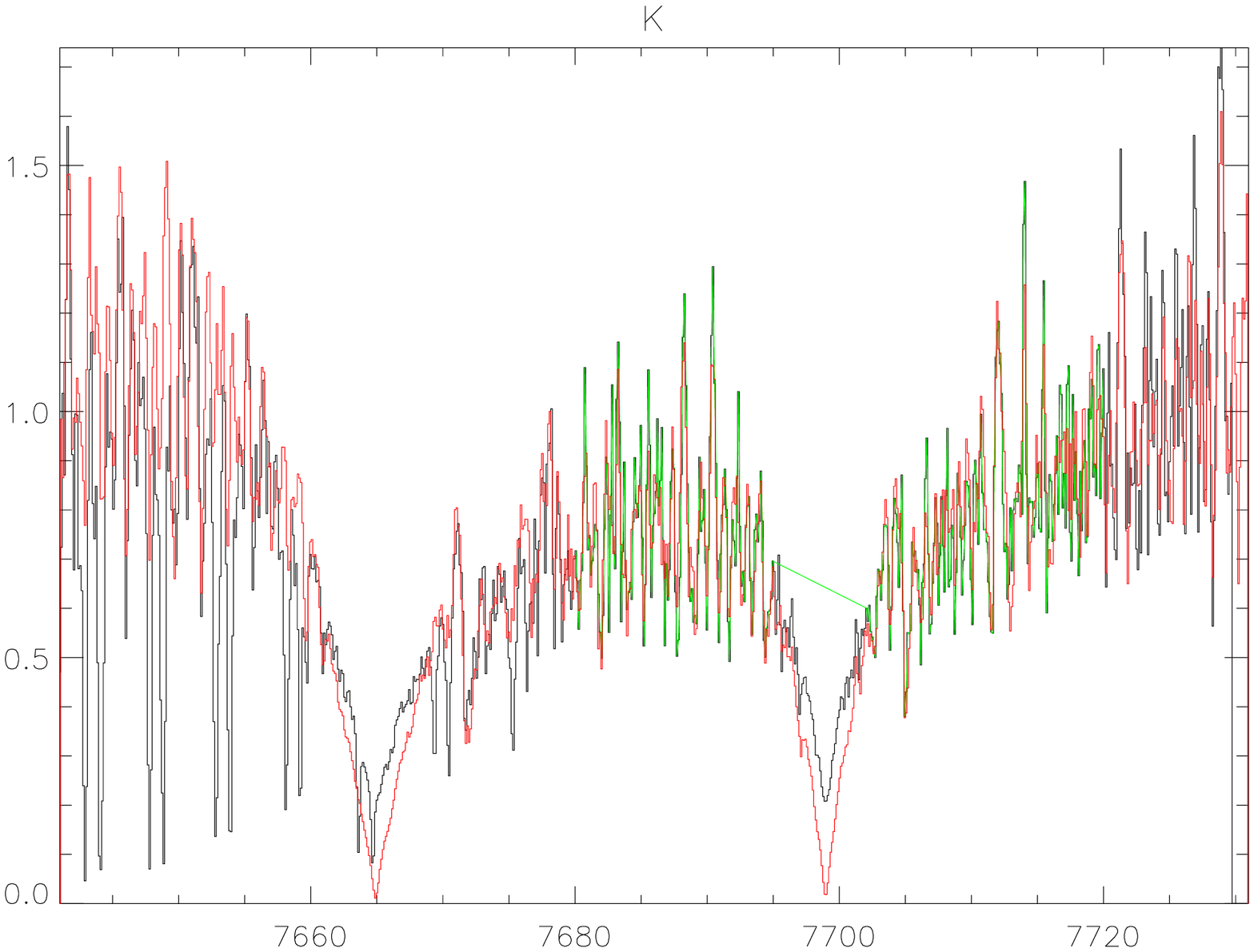}}
    \resizebox{.25\hsize}{!}{\includegraphics[]{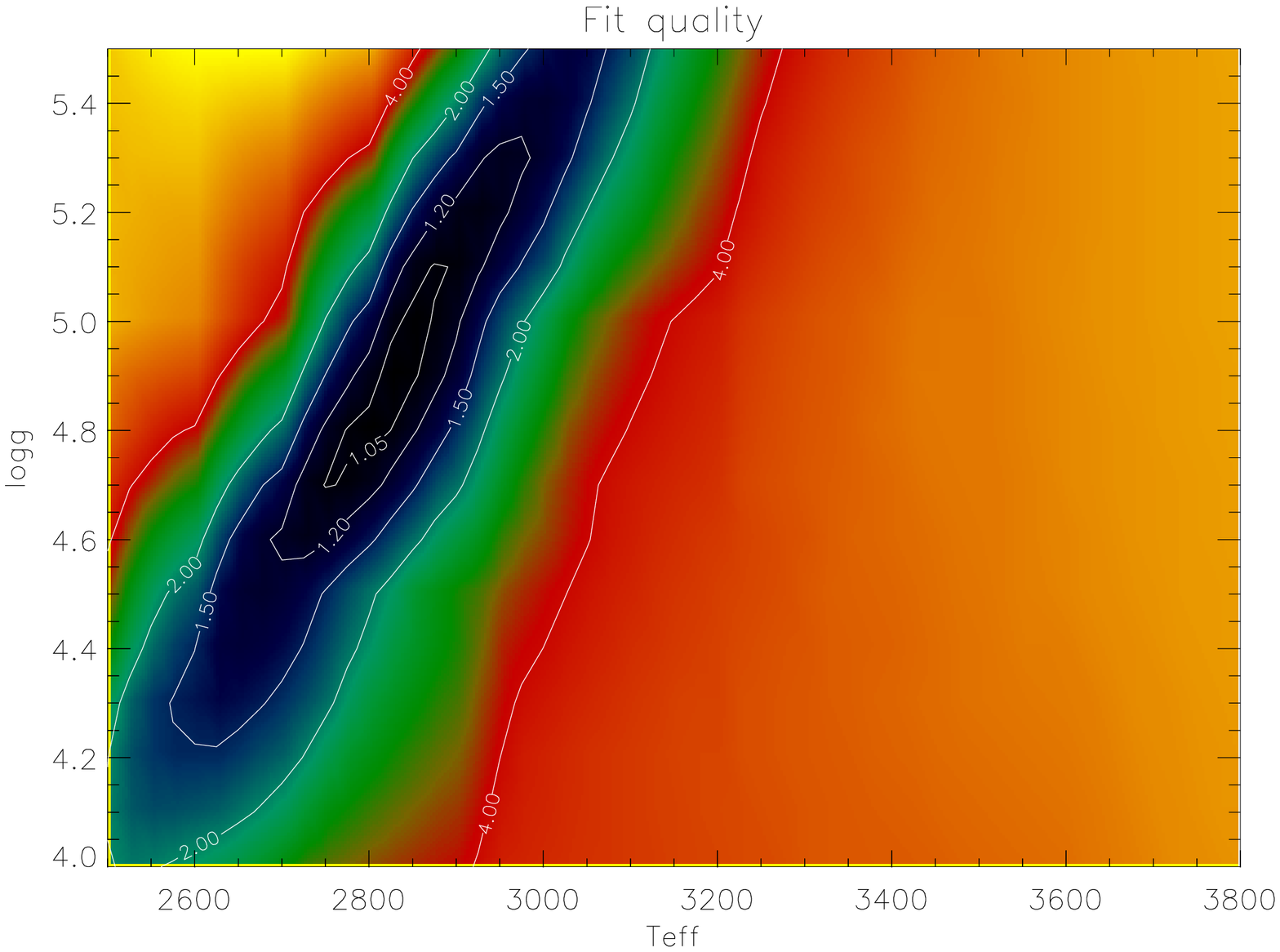}}
  }
  \caption{Fits of synthetic spectra to Na (left two panels) and K 
    (right two panels) resonance doublets of Gl\,406. For each case
    the spectral region is shown (black line: data; green symbols:
    data points used for fit; red line: fit). Right to the spectral
    region the fit quality is shown as a contour plot in $T_{\rm eff}$
    and log\,$g$, darker color means better fit. Both regions are
    sensitive to $T_{\rm eff}$ and log\,$g$ and yield consistent
    results.}
  \label{fig:logg}
\end{figure*}

\begin{figure*}
  \mbox{
    \resizebox{.25\hsize}{!}{\includegraphics[]{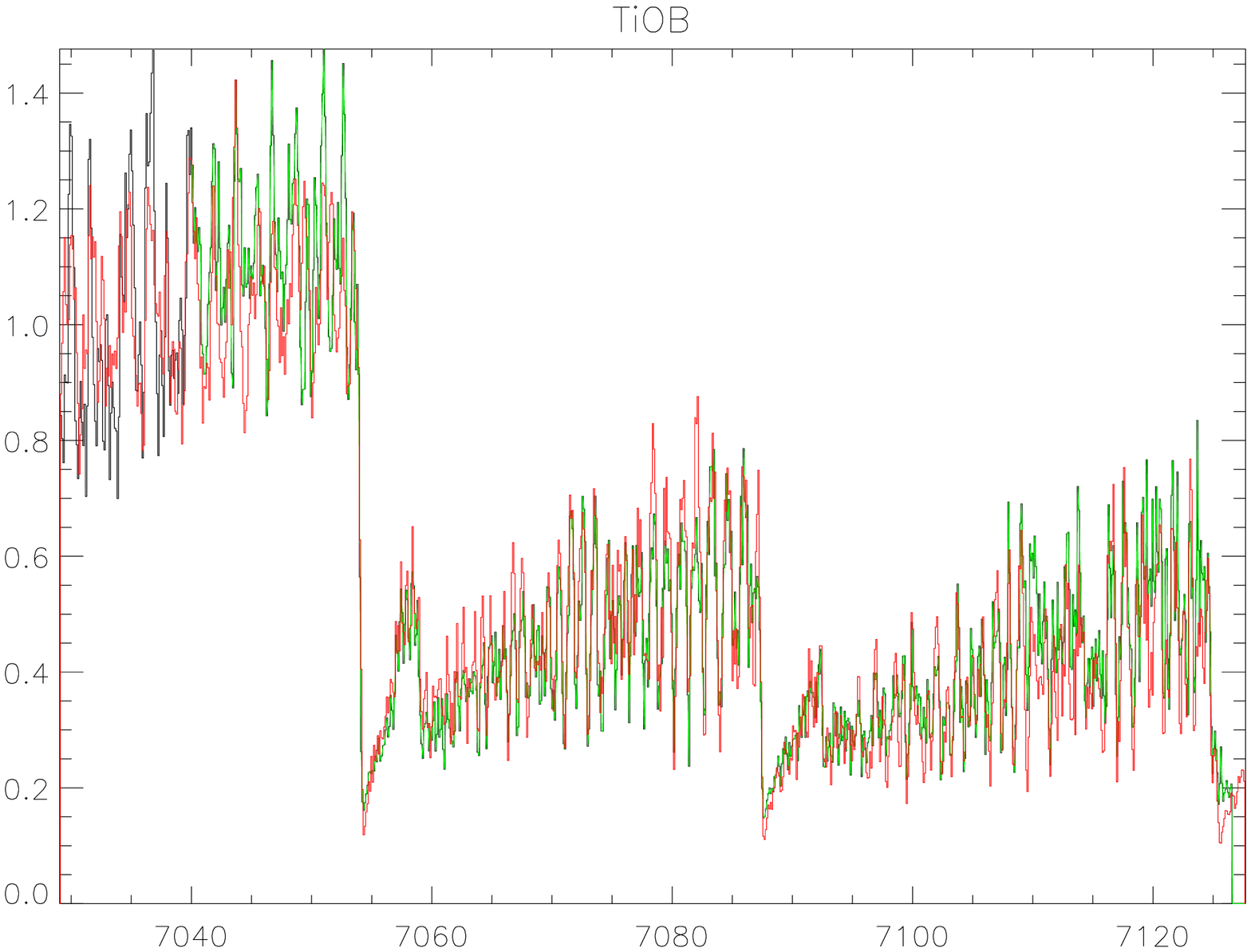}}
    \resizebox{.25\hsize}{!}{\includegraphics[]{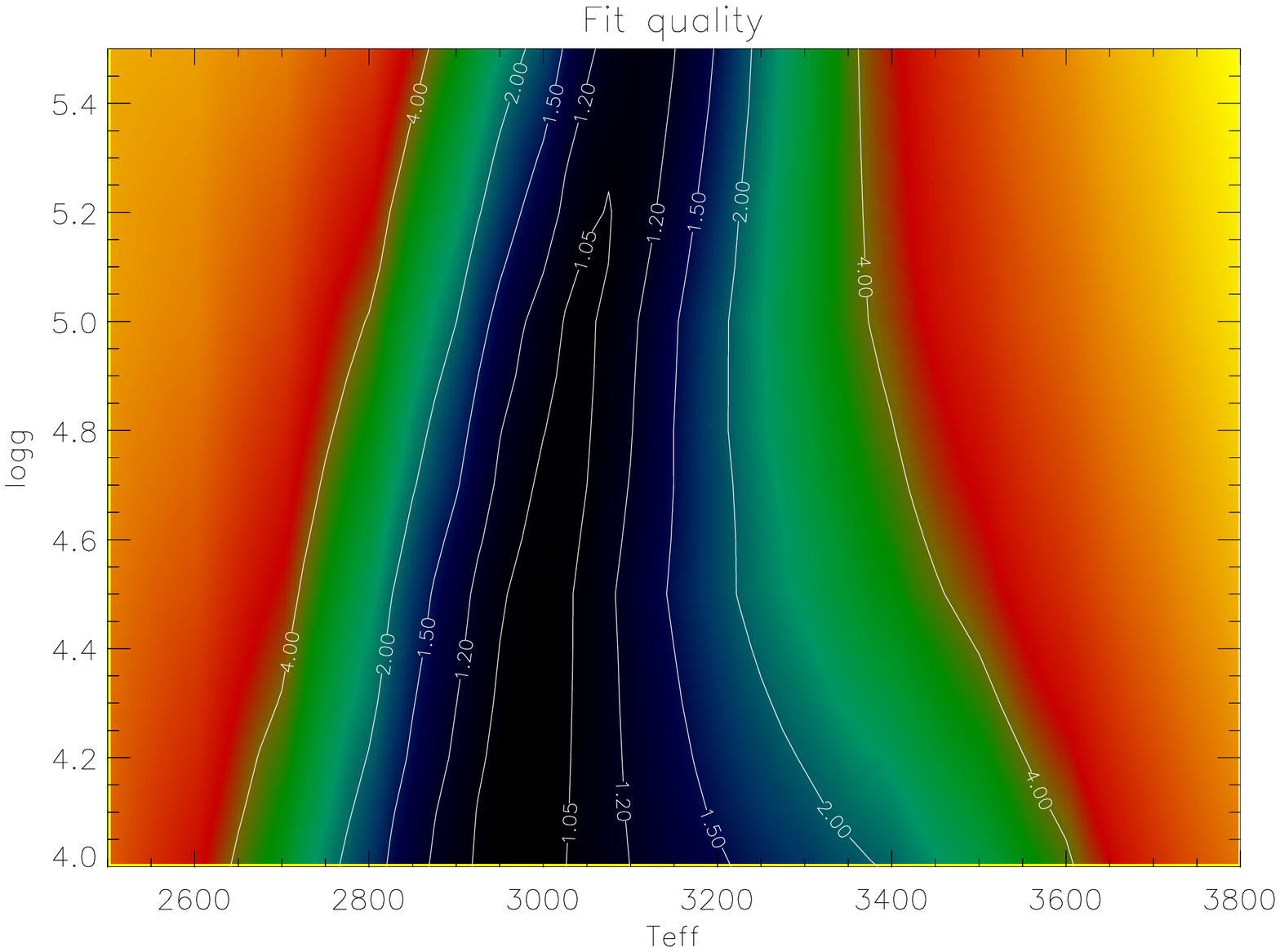}}
    \resizebox{.25\hsize}{!}{\includegraphics[]{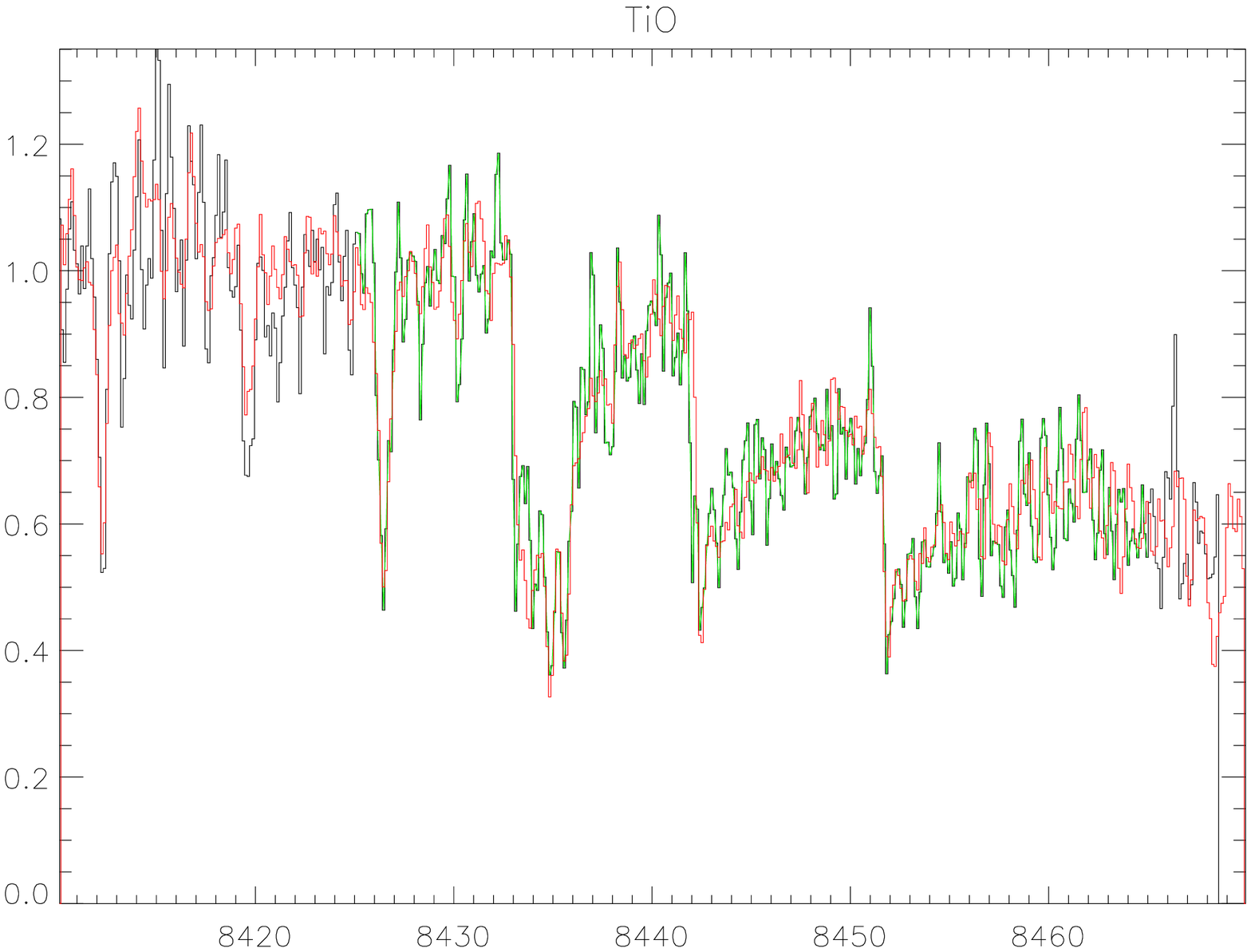}}
    \resizebox{.25\hsize}{!}{\includegraphics[]{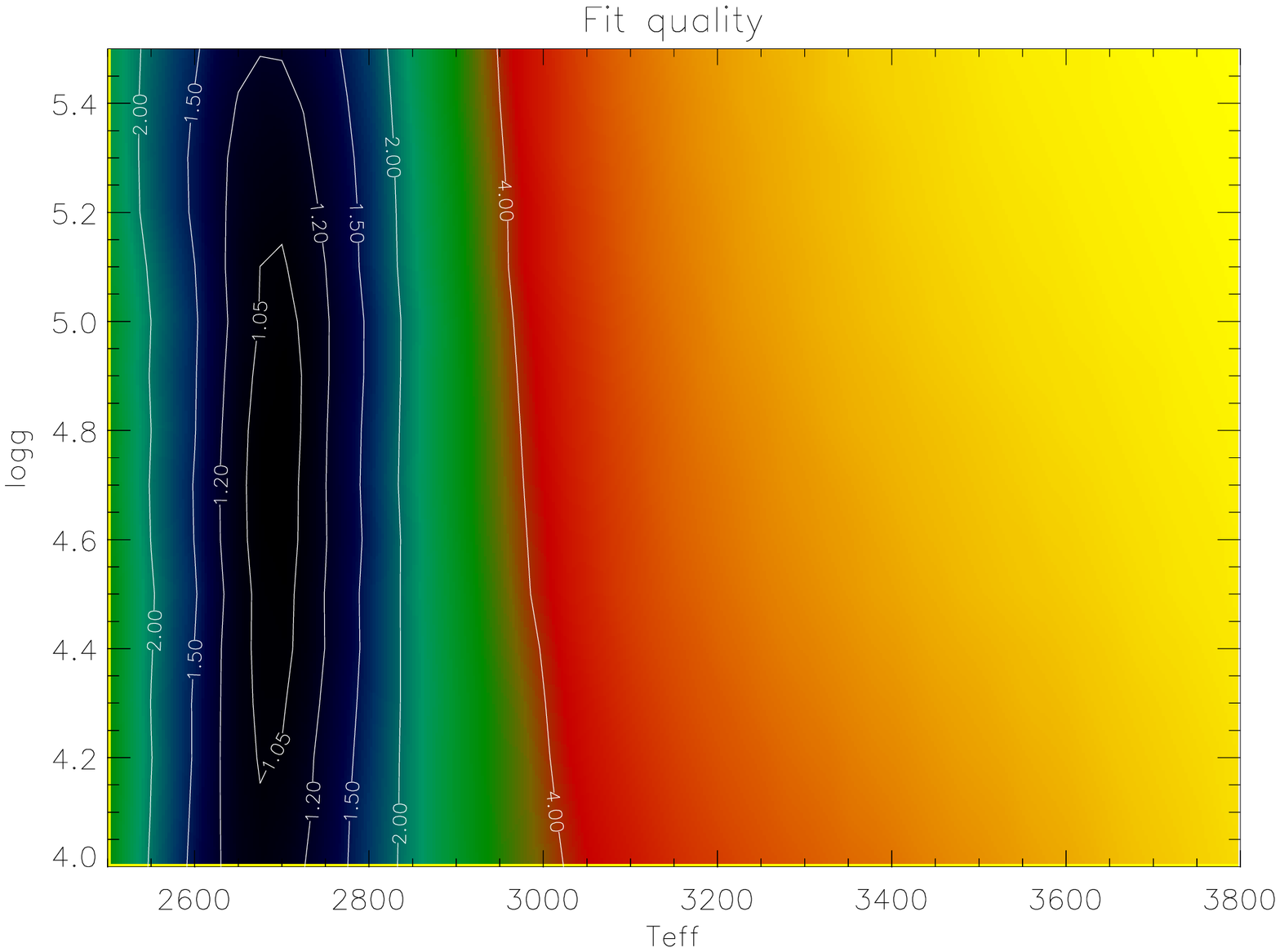}}
  }
  \caption{As Fig.\,\ref{fig:logg} but for the two TiO bands; $\gamma$-band 
    left, $\epsilon$-band right. Although good fit quality is achieved
    in both regions, the temperatures derived are inconsistent. Note
    that TiO does only marginally depend on gravity.}
  \label{fig:TiO}
\end{figure*}

A number of spectroscopic absorption features are very sensitive to
temperature and gravity. While temperature can also be derived from
low resolution spectra by investigating spectral energy distributions,
high spectral resolution is required to obtain the surface gravity
(low resolution spectra are only sensisitve to about one dex in
log\,$g$).

In Fig.\,\ref{fig:logg} the two gravity sensitive resonance doublets
of Na and K from a spectrum of Gl\,406 (CN\,Leo, M5.5V) are shown. For
each doublet the lines are shown with a model fit overplotted (see
caption).  Next to the spectra, the fit quality is displayed in a
contour plot as a function of $T_{\rm eff}$ and log\,$g$, darker color
implying a better fit. As can be seen in both contour plots, the lines
are extremely sensitive to temperature and gravity, but the best fit
is degenerate in $T_{\rm eff}$ and log\,$g$. Thus, it is possible to
determine a range in $T_{\rm eff}$ and log\,$g$ from Na and K lines in
high resolution spectra. However, to pin down a single solution in
temperature and gravity, it is necessary to obtain at least one
independent measurement in a region that shows a different
temperature/gravity dependence.

Such an independent measurement can be provided by molecular
absorption bands of TiO which are extremely sensitive to temperature,
but are virtually insensitive to gravity. Spectral regions with fits
to the same data as above and fit quality contour plots are shown in
Fig.\,\ref{fig:TiO} for two different bands of TiO ($\gamma$-band
left, $\epsilon$-band right).

The strategy is to derive the temperature from the TiO bands, and to
determine the surface gravity from Na/K with the temperature known
from TiO.  However, as can be seen in Fig.\,\ref{fig:TiO}, the two TiO
regions do not provide consistent temperatures. This does not only
rise the question which of the bands is more reliable, it implies that
some fundamental problem exists in the synthetic spectra of relatively
hot objects (note that at spectral type M5.5 dust formation is not a
problem). None of the temperatures derived from TiO should be trusted
without further investigation.

\section{Comparison to stars with known parameters}

To investigate the inconsistency of the model spectra, they can be
compared to spectra from stars with known temperature and gravity. The
probably most reliable values of $T_{\rm eff}$ and log\,$g$ come from
interferometric radius measurements which have been performed in some
M-dwarfs by S\'egransan et al.  (2003) and Dawson \& Robertis (2004).
Metallicities are available for some of them from a careful
investigation by Woolf \& Wallerstein (2005). Low mass stars with
measured radii and available high resolution spectra are listed in
Table\,\ref{tab:Objects}. For Proxima Cen a number of studies indicate
that its companions are at least of solar metallicity, and I use
[Fe/H]\,=\,0 yielding very good agreement in the $\gamma$-band.

\begin{table}
  \caption{\label{tab:observations} Calibrator stars for which temperature 
    and gravity are known from interferometric radius measurements}
  \label{tab:Objects}
  \begin{tabular}{llccc}
    \hline
    \hline
    \noalign{\smallskip}
    Object & Sp & $T_{\rm eff}$ [K] & logg & [Fe/H]\\
    \noalign{\smallskip}
    \hline
    \noalign{\smallskip}
    GJ 191 &  M1V   & $3570 \pm 156^{a}$    &    $4.96 \pm 0.13^{a}$  & $-0.86^{c}$\\
    GJ 205 &  M1.5V      & $3520 \pm 170^{a}$    &    $4.54 \pm 0.06^{a}$  & $-0.45^{d}$\\
    GJ 411 &  M2V        & $3570 \pm \,42^{a}$   &    $4.85 \pm 0.03^{b}$  & $-0.4\,^{c}$\\
    GJ 699 &  M4V   & $3134 \pm 102^{b}$    &    $5.04 \pm 0.10^{b}$  & $-0.75^{c}$\\
    GJ 551 &  M5.5V & $3042 \pm 117^{a}$    &    $5.20 \pm 0.23^{a}$  & $\ge0.0$\\ 
    \noalign{\smallskip}
    \hline
  \end{tabular}
  \begin{list}{}{}
  \item[$^{\rm a}$]S\'egransan et al. (2003)
  \item[$^{\rm b}$]Dawson \& Robertis (2004)
  \item[$^{\rm c}$]Woolf \& Wallerstein (2005)
  \item[$^{\rm d}$]Woolf \& Wallerstein (2005) report [Fe/H] = 0.21
    for $T_{\rm eff} = 3760$\,K; using $T_{\rm eff}$ and log\,$g$ from
    S\'egransan et al. (2003) yields the same TiO band strengths
    whith ${\rm [Fe/H]} = -0.45$
  \end{list}
\end{table}

\begin{figure*}
  \centering
  \mbox{
    \resizebox{.45\hsize}{!}{\includegraphics[]{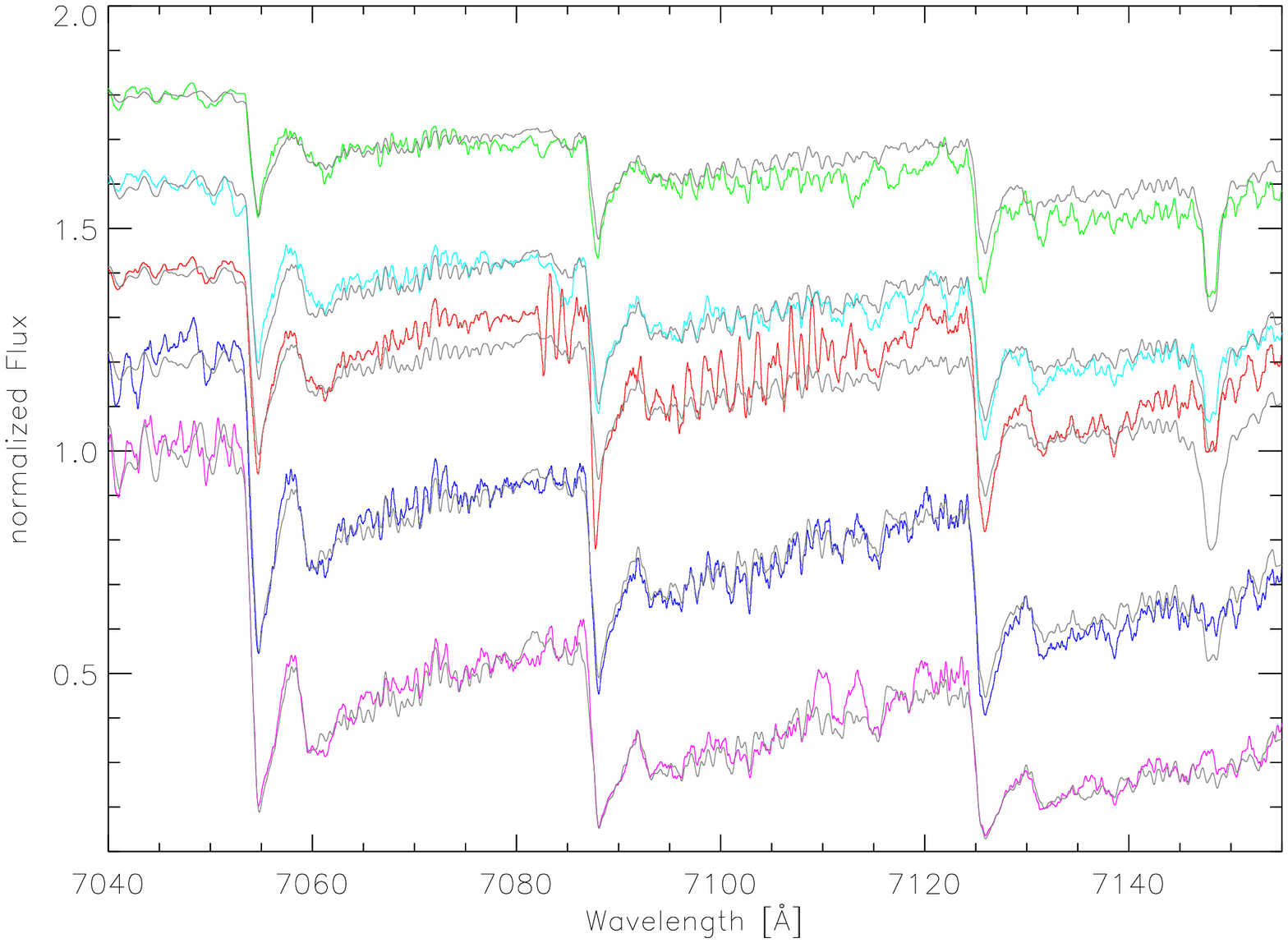}}
    \resizebox{.45\hsize}{!}{\includegraphics[]{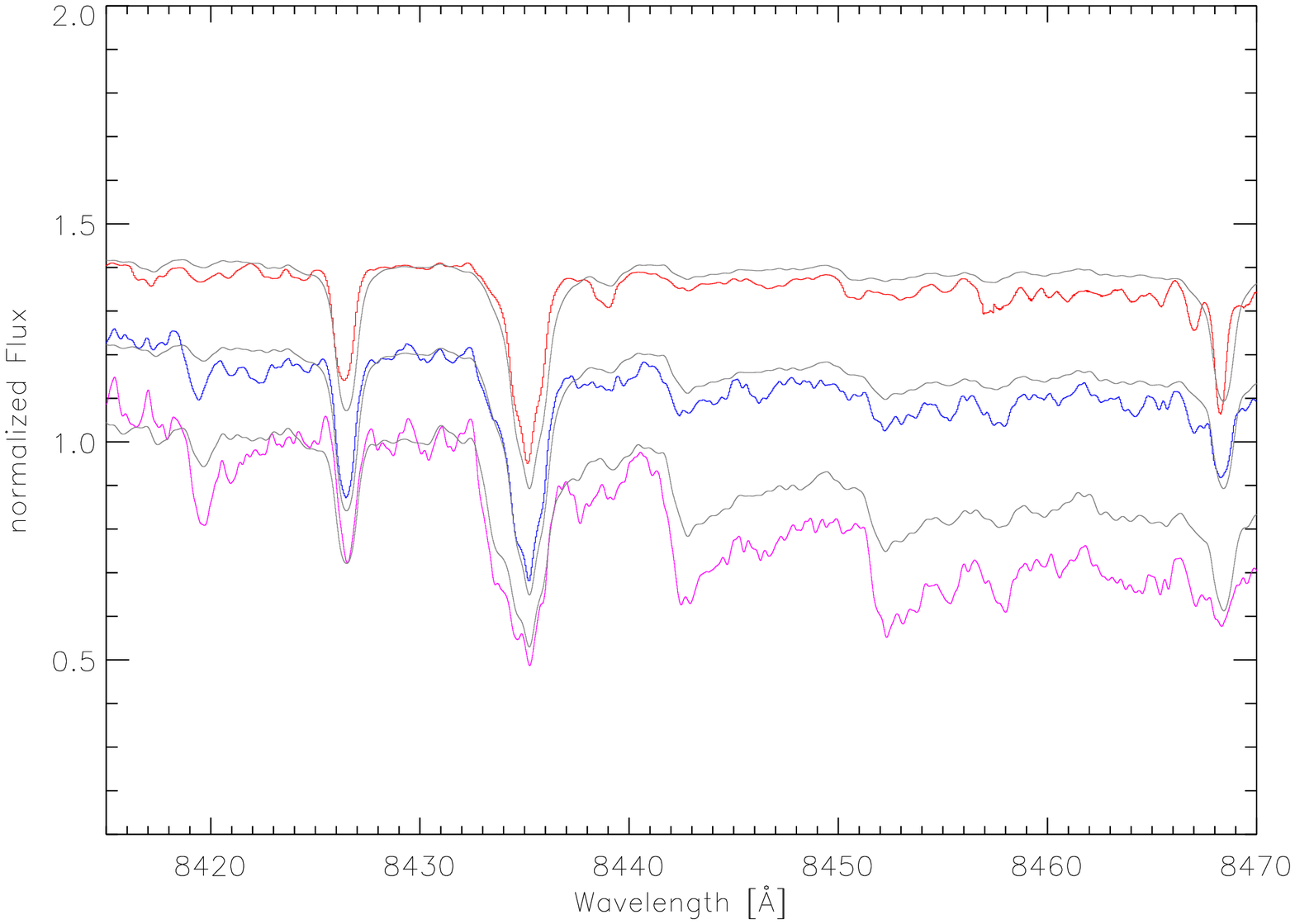}}
  }
  \caption{Spectra of stars with known $T_{\rm eff}$ and log\,$g$, overplotted 
    are models according to temperature and gravity. Left:
    $\gamma$-band; right: $\epsilon$-band. In the left panel from top
    to bottom: GJ\,191, GJ\,205, GJ\,411, GJ\,699 and GJ\,551. In the
    right panel only the three coolest stars are shown, the hotter two
    show no absorption in this region. Different spectra are plotted
    with an offset of 0.2.}
  \label{fig:calib}
\end{figure*}

The two TiO-bands of these stars are shown as colored lines in
Fig.\,\ref{fig:calib}.  Synthetic spectra according to their
temperature and surface gravity are overplotted as grey lines. The
$\gamma$-band appears stronger than the $\epsilon$-band; while the
former is visible in all targets, the latter is not detected in the
two hottest objects. GJ\,191 and GJ\,205 consequently are not shown in
the right panel of Fig.\,\ref{fig:calib}.

In all five calibrator stars, the model spectra provide an excellent
fit in the region of the $\gamma$-band (left panel of
Fig.\,\ref{fig:calib}). The model sequence reproduces the correct
absorption strength and its growth with decreasing effective
temperature. On the other hand, the $\epsilon$-band in the right panel
of Fig.\,\ref{fig:calib} does not fit the observations but
systematically underestimates the absorption strength in all three
spectra investigated. Thus, the temperature derived from the
$\gamma$-band can be expected to be very accurate, while the
temperature obtained from comparison to the $\epsilon$-system may be
systematically too low, since lowering the temperature would also
enhance the (underestimated) band strength and provide a better fit.

\begin{figure}
  \centering
  \mbox{
    \resizebox{.9\hsize}{!}{\includegraphics[]{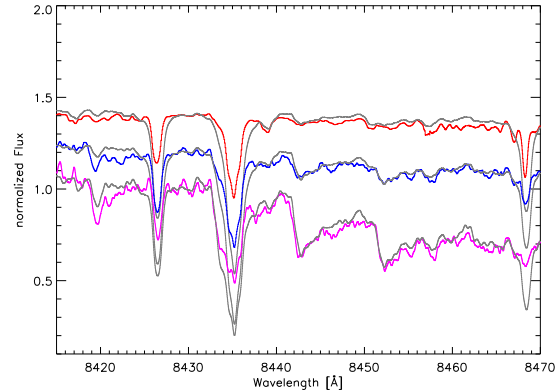}}
  }
  \caption{As right panel in Fig.\,\ref{fig:calib}, but with absorption features
    artificially enhanced by 70\%. The enhanced TiO band fits the data
    in all three cases.}
  \label{fig:calib_epsilon}
\end{figure}

In Fig.\,\ref{fig:calib_epsilon}, the $\epsilon$-band spectra are
shown as in the right panel of Fig.\,\ref{fig:calib}, but now the
absorption strength has been artificially enhanced by 70\% by simply
shifting the zero point (note that this also amplifies the atomic Ti
lines at for example 8435\,\AA, only the smooth features redward of
8440\,\AA\ are due to TiO). These modified models now provide
excellent fits in all three stars.  Together with the accurate
consistency achieved in the $\gamma$-band, this suggests that the
$\epsilon$-band is systematically underestimated in the models.

From this first estimate, it can be speculated what could be the
reason for the underestimated band strength in the $\epsilon$-system.
One reason can be that the current treatment of the partition function
of TiO is inappropriate (Allard, these proceedings). However, a
fundamental change of the TiO abundance probably would influence the
$\gamma$-band as well, but this one is in nice agreement with the
observations. A parameter that influences the $\epsilon$-band strength
only is the oscillator strength $f_{\rm el}$. As a first order
approximation, an oscillator strength 70\% higher than the one used
for the calculations would provide a TiO absorption strength that is
enhanced by roughly 70\% (the $\epsilon$ system is not saturated in
this regime).  Oscillator strengths for TiO from different ab initio
calculations and laboratory experiments together with the values
adopted for the currently available PHOENIX models are given in
Allard, Hauschildt \& Schwenke (2000). Their Table\,1 shows that for
the $\epsilon$-band $f_{\rm el}$ is not very well confined and a
slightly higher value would be in agreement with laboratory
experiments and calculations as well.

Of course, a different oscillator strength of the main opacity source
in very cool stars would have much more influence to the atmospheric
structure than just an enhancement of the absorption strength in one
molecular band.  Thus, the suggested enhancement of $f_{\rm
  el}(\epsilon)$ of 70\% can only be a first order guess that has to
be confirmed by self-consistently calculated atmospheres (and with the
correct treatment of the partition function).

\section{Implications for fundamental parameters}

\begin{figure}
  \mbox{
    \resizebox{\hsize}{!}{\includegraphics[]{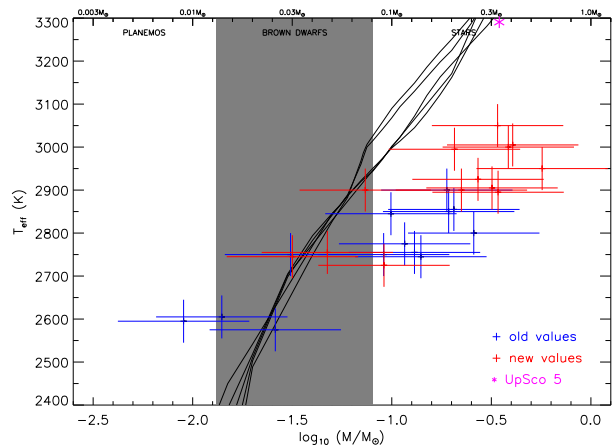}}
  }
  \caption{Temperature and mass for the members of the young association 
    UpSco (Mohanty et al. 2004b). Blue crosses show parameters derived
    from the original models, red crosses are derived using the
    adjusted TiO bands (see text).}
  \label{fig:Teff_Mass}
\end{figure}

Temperatures derived from model fits to the TiO $\epsilon$-band have
been used for example by Mohanty et al. (2004a, 2004b) to derive
fundamental parameters of young low mass objects. A reinvestigation of
$T_{\rm eff}$ from the modified $\epsilon$-band, which has been
artificially enhanced by 70\%, yields temperatures about 150--200\,K
higher than what was derived from the original models. Since
determinations of surface gravity in high resolution spectra employ
the effective temperature from TiO to dissolve the degeneracy in
$T_{\rm eff}$ and log\,$g$ (Sect.\,2), surface gravities are also
effected. A comparison to Fig.\,\ref{fig:logg} shows that a
temperature change of $\approx 200$\,K shifts the gravity to values
higher by roughly 0.3\,dex at 2700\,K (the correction does depend on
temperature and has to be individually calculated for each star).
Radii as well as masses consequently are also effected by a
temperature shift (cp. Basri, these proceedings).

In Fig.\,\ref{fig:Teff_Mass}, temperatures and masses of the sample
targets from Mohanty et al. (2004a, 2004b) are plotted (cp. Fig.\,3 in
Mohanty et al. 2004b). The original values are shown in blue, red
symbols display the new values according to a shift of 150\,K and the
corresponding gravity shift. Note that these new values are only
estimates from the qualitative changes discussed above, no proper
reanalysis has been done; Fig.\,\ref{fig:Teff_Mass} only displays the
approximate trends. Nevertheless, the discrepancy between the data
points and evolutionary tracks (shown as solid lines in
Fig.\,\ref{fig:Teff_Mass}) vanishes at very low masses.  In fact, the
lowest mass objects are shifted into the brown dwarf mass regime.  On
the other hand, the discrepancy becomes even larger for the hottest
objects, unphysically high masses of about one third of a solar mass
are estimated for objects of spectral type around M6 (cp.  Mohanty et
al.  2004b for details on the sample targets).  The recently
discovered spectroscopic binary UpScoCTIO~5 (Reiners et al.  2005) is
also plotted in Fig.\,\ref{fig:Teff_Mass}.  Although UpScoCTIO~5 shows
a significant discrepancy to the evolutionary models (which is hardly
visible in this figure), its dynamical mass of $M \ge
0.32$\,M$_{\odot}$ at a spectral type M4 is already lower than what is
derived for M6 objects using the high resolution fits. This clearly
shows that the far stronger disagreement between evolutionary models
and the corrected values in Fig.\,\ref{fig:Teff_Mass} must be due to
incorrect masses derived from the synthetic spectra.

\section{Summary}

Synthetic PHOENIX spectra have been compared to high resolution
spectra of stars with known atmospheric parameters. Absorption bands
of TiO are sensitive to temperature but insensitive to gravity. The
resonance doublets of Na and K are sensitive to both, $T_{\rm eff}$
and log\,$g$. Temperatures derived from the two TiO bands, which
belong to the $\gamma$ and $\epsilon$ system, differ by 150--200\,K.
The $\gamma$-band yields temperatures consistent with the known
parameters, $\epsilon$-band temperatures are systematically too low.
Enhancing the $\epsilon$-band by $~70\%$ makes this band fit to the
spectra. It is suggested that the oscillator strength employed for the
$\epsilon$-band is too low by about 70\%.

For the sample investigated by Mohanty et al. (2004a, 2004b), new
temperatures are estimated from modified models with artificially
enhanced band strengths, and the according gravities are estimated
from Na and K employing the corrected temperatures. The derived masses
are obviously too high suggesting a problem in the Na and K resonance
lines that are used to derive gravity and thus mass.

What can be the reason for the incorrect Na and K lines?  The Na lines
are embedded in the TiO-$\epsilon$ system which is probably too weak
as shown above. Stronger TiO absorption will effect the Na lines but
is not expected to have much influence on the K line.  It is important
to note that both doublets show the same (wrong) behaviour, which
makes it probable that only one mechanism is responsible for the wrong
line shape in both systems. Burrows \& Volobuyev (2003) and Allard et
al.  (2003) present an improved model on line shape theory for the Na
and K resonance lines. Their improvements have not yet been
incorporated into the PHOENIX models and should be implemented soon
(Allard, these proceedings).

These major improvements, namely more accurate $\epsilon$-band
oscillator strengths and improved Na and K line shape theory, are the
most promising candidates to provide better Na and K lines and to
yield correct temperatures, gravities, and hence masses. As soon as
new models are available, it is necessary to compare the new Na and K
lines to the stellar spectra as done here. Such a comparison is an
essential first step in order to correctly interpret any
inconsistencies -- and also the agreements -- between observations and
evolutionary models.

\acknowledgements AR has received research funding from the European
Commission's Sixth Framework Programme as an Outgoing International
Fellow (MOIF-CT-2004-002544).

\end{document}